\documentclass[runningheads]{llncs}

\usepackage{cite}
\usepackage{amsmath,amssymb,amsfonts}
\usepackage{algpseudocode}
\usepackage{graphicx}
\usepackage{textcomp}
\usepackage{xcolor}
\usepackage{hyperref}
\usepackage{multirow}
\usepackage{caption}
\usepackage{adjustbox}
\usepackage{subcaption}

\hypersetup{
    colorlinks=true,
    linkcolor=blue,
    filecolor=magenta,      
    urlcolor=cyan,
    pdftitle={Enabling Property Graph Analysis in Arachne},
    pdfpagemode=FullScreen,
}
\usepackage[linesnumbered,ruled]{algorithm2e}

\begin{document}

\title{Profile of Vulnerability Remediations in Dependencies Using Graph Analysis}

\author{Fernando Vera\inst{1} \and
Palina Pauliuchenka\inst{1} \and
Ethan Oh\inst{1} \and
Bai Chien Kao\inst{2} \and
Louis DiValentin\inst{2} \and
David A. Bader\inst{1}}

\institute{Department of Data Science, \\ New Jersey Institute of Technology, Newark, NJ, USA \email{\{fv54,pp272,eo238,bader\}@njit.edu} \and
Accenture PLC \email{\{bai.chien.kao, louis.divalentin\}@accenture.com}}

\maketitle
\begin{abstract}
This research introduces graph analysis methods and a modified Graph Attention Convolutional Neural Network (GAT) to the critical challenge of open source package vulnerability remediation by analyzing control flow graphs to profile breaking changes in applications occurring from dependency upgrades intended to remediate vulnerabilities. Our approach uniquely applies node centrality metrics—degree, norm, and closeness centrality—to the GAT model, enabling a detailed examination of package code interactions with a focus on identifying and understanding vulnerable nodes, and when dependency package upgrades will interfere with application workflow. The study's application on a varied dataset reveals an unexpected limited inter-connectivity of vulnerabilities in core code, thus challenging established notions in software security. The results demonstrate the effectiveness of the enhanced GAT model in offering nuanced insights into the relational dynamics of code vulnerabilities, proving its potential in advancing cybersecurity measures. This approach not only aids in the strategic mitigation of vulnerabilities but also lays the groundwork for the development of sophisticated, sustainable monitoring systems for the evaluation of work effort for vulnerability remediation resulting from open source software. The insights gained from this study mark a significant advancement in the field of package vulnerability analysis and cybersecurity.

\end{abstract}

\section{Introduction}

Preventing and analyzing vulnerabilities is crucial for averting cyber attacks and protecting programs and components \cite{yan2020review}. In the face of the increasing complexity and abstraction via dependencies and version management, the likelihood of vulnerabilities residing within application code in this dynamic, continuously updating world is heightened \cite{liu2022review}. It's clear that during code development, it is not only essential to identify and address the weakest link or the most apparent vulnerability, but also to pursue continuous solutions with each code update \cite{liu2022recent}. To effectively protect yourself from cyber attacks, three steps are proposed: 1) know the vulnerabilities, 2) understand their impact on the application, and 3) act on them in depth.

%%%%%based in fnctions->codeql y repositoy

Vulnerabilities remediation in open source packages is an essential process for the open source community and the security of many users. This process consists of identifying vulnerabilities, reporting them, and analyzing their impact \cite{alfadel2023empirical}. Open source maintainers then triage those vulnerabilities and bugs and prioritize updates, with each update for the affected package fixing affected functions and removing bugs. The new package is then updated to the community without the targeted vulnerability.

%%But, will this package upgrade (which is intended to fix vulnerability) break my application.

Various tools are available to facilitate the discovery of vulnerabilities in dependencies. CVEfixes is a repository of code that can be used to replicate the comprehensive database of vulnerabilities curated from Common Vulnerabilities and Exposures (CVE) records in the public U.S. National Vulnerability Database (NVD). Static and dynamic code analysis capabilities can be used to used to discover new vulnerabilities. Additional code analysis techniques can be used beyond static analysis as well. There are a variety of techniques used to perform vulnerability analysis on control flow graphs \cite{sparks2007automated} for the application. CodeQL \cite{szabo2023incrementalizing} is a powerful tool used by the GitHub community to identify and modify functions during their development.

Some scanning techniques and vendors have emerged to address vulnerabilities residing in the package dependencies. Generally these solutions perform a Software Composition Analysis (SCA) \cite{imtiaz2021comparative} technique in order to create a Software Bill of Materials (SBOM). A typical SBOM lists (to varying degrees of depth) the specific versions of dependencies used to compile the software \cite{xia2023empirical}. This dependency version list can be correlated to existing vulnerability repositories to find the exact vulnerabilities present in each specific dependency used to compile the software. Those specific dependencies with vulnerabilities can be targeted for upgrade to newer versions where those vulnerabilities are fixed by changing the version number and recompiling the code with the newer version. Attempting to migrate to a newer version sometimes causes errors however, and the application will refuse to compile, or have erroneous functionality when run.

%%%based in knowlege graph->history
Understanding the impact of changes between dependency versions, along with the functional interconnections and network structures is a task of that can be used to estimate the likelihood of a specific package upgrade causing breaking changes in application code.  Given that all applications are distinct and highly adaptable, there is a need for tools that generalize across different applications and can provide insights into the complexity of the upgrade. For this reason, we turn to knowledge graphs.\cite{sikos2023cybersecurity,jia2018practical} Constructing a graph enables us to comprehend the interactions between functions within the code. In the context of this research, the aim is to represent functions in the code as entities in a graph space. Constructing a Control Flow graph of interactions between functions in the code allows examination of a representation of the application code and dependencies in the graph space, with the caller-callee relationship serving as the connecting link \cite{zhu2021constructing}. This approach allows us to observe and analyze interactions and relationships between functions. We propose that gaining a relational understanding of code can assist in identifying the impact of changes in dependencies on the workflow of the application and the downstream effects, allowing individuals to move closer to full automation of package upgrades with minimal impacts to functionality. This can directly translate to improved security by allowing automated processes to upgrade vulnerable package versions and close vulnerabilities when upgrades cause a minimal impact to the functionality of the application. 

%%%topologia.
Understanding the impact of changed function nodes between two versions of an application using a graph is necessitates a profound comprehension of the network's topology and its connectivity attributes \cite{zomorodian2009computational}. This knowledge is proving indispensable in identifying the significance of the affected nodes to the overall operation of the network and in making informed interventions on the impact of an upgrade decision \cite{arulselvan2009detecting}. This process is involving an intricate analysis of their centrality, density, connected components, degree associativity coefficient, and the pathways traversing through them. Such comprehensive evaluations are facilitating strategic decision-making for optimally mitigating vulnerabilities, thereby ensuring a more robust and secure network architecture. This aspect is particularly critical in software, where functions can exhibit complex nested dependencies and feedback loops.

\section{Background and Related Work}
In the realm of package function analysis, traditional methods have primarily revolved around static and dynamic analysis techniques. Static analysis, while useful for a broad overview, often lacks the context-specific insights garnered from dynamic runtime analysis \cite{park2019estimating}. Additionally, these techniques typically deal with implementations of functions within the application, and not vulnerabilities emerging from the dependencies of the application.
Some scanning techniques and vendors have emerged to address vulnerabilities residing in the package dependencies. Dependency Vulnerabilities are remediated in this paradigm by upgrading the package version to a version in which the vulnerability is closed and recompiling the code. If the version of the package with the remediation causes the software to no longer compile or fail unit tests, then additional work will be necessary to update to application code to be compatible with the remediated package version. 
Understanding the effects of upgrading from the current package version to a remediated package version is currently an opaque process. 

Control flow graph analysis is a useful tool for understanding software behavior, allowing developers to easily visualize the flow of execution within a program. By reducing abstract and complex code into manageable graphics, it provides information about the structure of the program and its possible vulnerabilities and affected functions. This analysis is crucial to optimize code performance and ensure strong cybersecurity measures. In our case, we replace the control flow with a knowledge graph related to the use of algorithms and measures focused on large-scale analysis.

Graph Attention Networks (GAT) have emerged as a promising tool in similar analytical contexts. Originally conceptualized by Veličković et al.\cite{velivckovic2017graph}, GATs leverage the attention mechanism to provide node-specific contextual insights, enhancing the accuracy of feature representation in graph-structured data. This attribute of GATs has been effectively utilized in various domains, including but not limited to, bioinformatics, social network analysis, and natural language processing \cite{xia2023coupled}. In package function analysis, employing GAT presents an innovative opportunity to address the existing limitations of traditional methods. It offers a scalable approach to analyze the intricate interrelationships and dependencies among package functions represented as nodes in a graph. This method aligns with the recent trends in utilizing deep learning techniques for software engineering challenges, as documented in several contemporary studies \cite{giray2021software}.

\section{Methodology}
This section is outlining key definitions and delving into the critical elements that underpin our methodological approach. A portion of our research is the deployment of a modified Graph Attention Network (GAT), an advanced model that is being specifically designed to analyze feature code interactions and study the importance of changed functions represented as nodes in the graph.

\subsection{Modified Graph Attention Neural Network}
The adaptation of the GAT model is motivated by the lack of current visibility into the effects of upgrading package versions, which directly affects developers ability to stay on top of package vulnerabilities originating from dependencies. 
Unlike traditional GAT models based solely on norm \cite{velivckovic2017graph}, our modified version incorporates essential node centrality metrics (degree centrality, norm, and closeness) to evaluate the importance and interrelationships of various functions within the code, with a particular focus on identifying and evaluating vulnerable nodes. This enhancement increases the ability of GATs to provide a more refined analysis of the relational dynamics within the code, offering a comprehensive perspective on the impact of changes in dependency versions and the potential risk of breakage resulting from changed code.

\subsection*{Rationale for Modification}
Utilizing the Knowledge Graph for network analysis, particularly emphasizing on graph attention Neural Network (GAT) as defined in the work of Veličković et al. \cite{velivckovic2017graph}, is enabling us to train extensive datasets for the purpose of evaluating and quantifying the significance of nodes within a network. This base model is facilitating accurate detection of nodes based on both transductive methods (such as Cora, Citeseer, Pubmed) and inductive methodologies. Presented as a convolutional-like neural network operating on knowledge graph-structured data, this model assigns an importance metric to nodes within their neighborhoods.

The input to our layer consists of a set of node features, $h = \{\mathbf{h}_1, \mathbf{h}_2, \ldots, \mathbf{h}_N\}$, with $\mathbf{h}_i \in \mathbb{R}^F$, where $N$ denotes the number of nodes and $F$ the number of features per node. This layer is producing a new set of node features, $h' = \{\mathbf{h}'_1, \mathbf{h}'_2, \ldots, \mathbf{h}'_N\}$, with a potentially different cardinality $F'$. A shared linear transformation, parameterized by a weight matrix $\mathbf{W} \in \mathbb{R}^{F' \times F}$, is applied to each node. The GAT then performs a self-attention mechanism $a$, yielding attention coefficients $e_{ij}$ as:

\begin{equation}
e_{ij} = a(\mathbf{W}\mathbf{h}_i, \mathbf{W}\mathbf{h}_j) \label{eq:1}
\end{equation}

These coefficients indicate the importance of node $j$'s features to node $i$. The GAT model is assessing the first-order neighbors of $i$, necessitating a normalization of coefficients across all node features $j$ using the softmax function:

\begin{equation}
\alpha_{ij} = \text{softmax}_j(e_{ij}) = \frac{\exp(e_{ij})}{\sum_{k \in \mathcal{N}_i} \exp(e_{ik})} \label{eq:2}
\end{equation}

it is calculated the coefficient most relevant  of each feture of node using the norm. It identified as $\alpha_{ij}$. Replacing equation (\ref{eq:1}) in equation (\ref{eq:2})

\begin{equation}
\alpha_{ij} =\frac{\exp\left( \text{LeakyReLU}\left(a^\top [\mathbf{W}{h}_i |\mathbf{W}{h}j] \right) \right)}{\sum{k \in \mathcal{N}(i)} \exp\left( \text{LeakyReLU}\left( {a}^\top [\mathbf{W}{h}_i |\mathbf{W}{h}_k] \right) \right)}
\end{equation}

In line with the original GAT model \cite{velivckovic2017graph}, with the attention mechanism $a$ as a single-layer feedforward neural network, we define a parameterized weight matrix $\mathbf{a} \in \mathbb{R}^{2F'}$ and apply the LeakyReLU nonlinearity (negative input slope of 0.2). $a^\top$ is a learnable weight vector in the single-layer feedforward neural network which constitutes the attention mechanism. $\mathcal{N}(i)$ denotes the neighborhood of node $i$ in the graph. $\|\|$ represents the concatenation operation.

\begin{equation}
e_{ij} = a(\mathbf{W}\mathbf{h}_i, \mathbf{W}\mathbf{h}_j)
\end{equation}

This process calculates the most relevant coefficient for each feature of a node using the norm, identified as $\alpha_{ij}$. To enhance the focus and applicability of these coefficients, particularly in code analysis, we are introducing a modification $\beta_{ij}$ to highlight the robustness and criticality of all nodes. This modification encompasses functions such as degree centrality, the norm, and closeness centrality metrics, relevant to nodes $i$ and $j$. Consequently, the revised attention coefficient, $\alpha'_{ij}$, is defined as:

\begin{equation}
\alpha'_{ij} = \alpha_{ij} \cdot \beta_{ij}
\end{equation}

This novel approach is proving instrumental in evaluating the importance of specific functions within their neighborhoods, offering insights into the criticality of an affected function in a code structure. The modified model enables the generation of a normalized score ranging from $0$ to $1$, which reflects the importance of nodes based on metrics such as degree, norm, and centrality. This scoring system facilitates a more nuanced understanding of node significance within the graph structure, particularly in the context of package code analysis.

\subsection{Mapping and Mitigating Vulnerability in a Code Development}
\begin{figure*}
    \centerline{\includegraphics[width=1\textwidth]{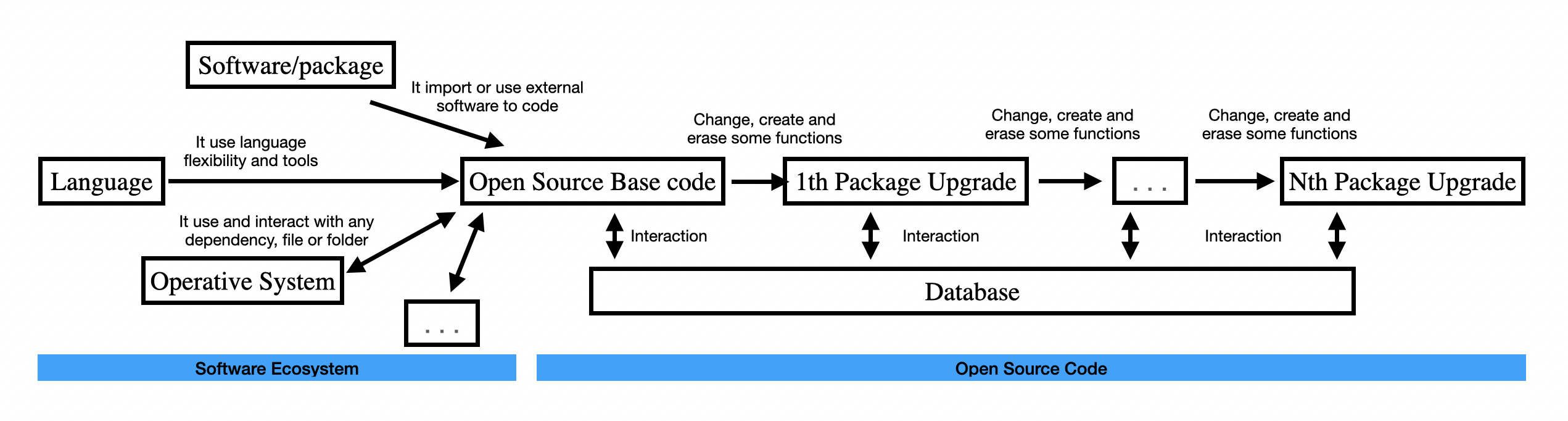}}
    \caption{In this figure, we are providing a conceptual description of the actors considered behind the open source. We are taking into account a generic software ecosystem of code from key elements such as the operating system, language, and software, packages or others that enable the code to function and connect with the real world. A code under development typically begins with the Open Source Base code (primary functions) that is interacting with the repository, the operating system, and other components of the ecosystem. This base code is undergoing updates $Nth$ times to enhance its functionality. It is crucial to emphasize that within each component, there could exist certain flaws which might potentially manifest as vulnerabilities within the code.}
    \label{fig:graph-code}
\end{figure*}

Vulnerabilities are recorded in the NVD repositories from various sources during the development process. These can be inherent from the base of the code (fig.\ref{fig:graph-code}) to the programming language used, integrated through specific packages or software, or due to operating system flaws\cite{10.1145/3379597.3387501,goseva2015capability}. These vulnerabilities are remediated by standardizing community-driven coding practices\cite{seacord_2014}, like SAST and DAST scanning, and bug reporting. Nevertheless, vulnerability mapping is continuously evolving, with new vulnerabilities emerging or being discovered over time, as well as new tools and methods for finding them\cite{9802103,varela2023feature}. Thus, it is proving beneficial to maintain an updated, dynamic, and comprehensive connection map, which serves as a guide to pinpoint the source and address these vulnerabilities swiftly and precisely.

From the foundational layer of code, the language in which it is being written is emerging\cite{alfadel2023empirical}, along with the environment and configurations is more correct and the built-in packages\cite{alanazi2022scada}. From this foundation, the core code is developing, forming the heart of the program with its principal functions. An initial version of a code is committed, and subsequent new functionality and features are added along with errors being resolved with updates. Each update introduces a new layer of functions and packages that can improve the structure of the code but also can also add complexity (fig. \ref{fig:graph-code}). A new update may be integrating previous functions into the core (at the discretion of each software developer) or may be leaving them as branches outside the core. Over time, vulnerabilities can appear in any new update or may be stem all the way back to the base code. A vulnerable code fragment, if reused across updates, can extend its reach to critical branches, potentially impacting the entire application's core structure.Understanding the network of functions within the code is crucial to mitigate such risks and ensuring that updates do not inadvertently compromise the integrity of the application.

\subsection{Knowledge Graph in Cybersecurity}
According to Hogan \cite{Hogan_2021}, a debate is ongoing regarding the precise definition of a Knowledge Graph, yet consensus exists about its remarkably high adaptability. In the context of this discussion, the knowledge graph $G$ is defined as $G = (E, R, S)$, where $E$ represents the entity (node), $R$ symbolizes the relationship (edge), and $S$ denotes relationship facts (node-node relationship). A triplet constitutes a typical form of knowledge representation within this framework. Entities, serving as foundational elements of the Knowledge Graph, encompass a wide range of classifications, such as collections, categories, object types, and thing categories (e.g., domain, host, etc.). Relationships interconnect these entities to formulate the graph's structure, while attributes encapsulate features and parameters, exemplified by entities like google.com, windows, and similar.

To construct a dataset, studying the relationships existing between functions through their callee or caller interactions is necessary. A database is generated considering the node entities $E_i$, their relations $R_i$, and their relation $S_i$, which can be a call or a caller. This approach enables us to generate a knowledge graph containing the identified entities.

The data structure is providing critical information, encompassing the function's path, its name, and whether it has been modified in the latest update. Additionally, it indicates whether the function is vulnerable and specifies its role as either a callee or a caller.

\subsection{Building the Dataset}

To build a comprehensive dataset, we compare subsequent versions of code where upgraded package versions are intended to remediate vulnerabilities in the open source packages \cite{ alanazi2022scada}. We first use the open source software Syft to generate the software bill of materials (SBOM) information for the current version of the target application source code. Curl is used to fetch vulnerability data for individual package versions in the SBOM from the Open Source Vulnerabilities (OSV) database, which offers an accessible query interface for all known dependency vulnerabilities. This information is mapped to the SBOM to identify the existing vulnerabilities in the dependencies. If a package with an existing vulnerability is detected, we search for the updated version of the package and clone it. Using the CodeQL, a semantic code analysis engine, we model the different versions of the code and construct a control flow graph of the execution paths for the repository using the vulnerable package and the fixed package. Tree-sitter is applied to perform incremental analysis, construct and maintain a syntax tree, and build the dataset. The impact of a single package upgrade can be measured by comparing the control flow graphs of the application before and after the upgrade. The updated version control flow graph is used as the base for the knowledge graph, and changed or affected nodes are marked in the knowledge graph, along with nodes that cause errors in compilation of the code. Multiple upgrades are performed split into two sets for each each repository, one set consisting of graphs introducing errors that result in code breakage, and another set of graphs not impacting code functionality.

In this research, a preliminary approach is being used, initially focusing on three application source repositories; The first case in a code base comprising 9621 features with 27 vulnerabilities in its broken update; the second case with 19,569 functions and 3 vulnerabilities. Subsequently, in Cases 3 (15908 functions and 6 vulnerabilities) and 4 (16095 functions and no vulnerabilities), an application is observed with a similar order of magnitude but different dependencies regarding the base functions.

\section{Analysis}
In this section, we discuss the methods of data analysis and describe the metrics we will use. Once the graph structure has been built, we can utilize tools for small-scale analysis like NetworkX in Python \cite{platt2019network} or for large-scale projects: Arachne \cite{rodriguez2022arachne}. In this case, NetworkX is used exclusively to facilitate calculation and analysis, given the volume of data (less than 20000 nodes). This analysis adopts a two-pronged approach. Firstly, it involves an exploratory examination of graph connectivity, focusing on the distribution and other metrics that provide insights into the difference between package upgrades that break dataflow vs. those that do not and quantitative metrics on the differences in interconnections within the code. This includes evaluating the structural design and robustness of the network, ensuring that each functional node and its connections contribute to the overall integrity and efficiency of the system. Secondly, we analyze the modified GAT scores to give us a normalized measure the interaction of vulnerabilities as nodes within the graph. The model gives us an advanced mechanism to measure the importance of nodes within a network. This approach allows for a targeted strategy for classifying and assessing vulnerabilities, simplifying the assessment of remediation effectiveness. The practical benefits of this approach are many and include prioritizing development efforts, improving software integrity, formulating strategic planning initiatives, and deepening knowledge about software architecture. This comprehensive strategy not only improves the immediate security posture of software systems, but also lays the foundation for long-term sustainable software development and maintenance practices.

%%% By integrating both connectivity analysis and vulnerability assessment, the study aims to provide a holistic understanding of the network's architecture and its resilience against potential threats.

\subsection{Graph analysis}
It is crucial when analyzing the knowledge graph created with the data set to measure the network and understand the meaning and relevance of the entities and relationships within it \cite{1690659}. It is necessary to quantify the connectivity, the paths and their strength through the centrality analysis of each objective element.

\subsection*{Preliminar Graph Analysis}

We are constructing a knowledge graph, denoted as $G = (\hat{E}, \hat{R}, \hat{S})$, where $\hat{E} = V$ signifies the entities or nodes of the graph. Our primary focus is on the callee-caller relationship between these entities, which we are considering as a directional edge. This approach enables us to simplify the triplet with the tuple $(\hat{R}, \hat{S})=E$ as the set of of edges $E$ into the graph structure $G = (V, E)$ \cite{1690659}. This representation allows us to prepare a knowledge graph in relation to the GAT model \cite{velivckovic2017graph}.

This knowledge graph model is characterized by a dynamic representation of functions, where the total number of vertices is indicated by $n$, and the edges, denoted by $m$, are distinctly directed. Each edge  $e \in E$ in this graph is assigned a specific weight, denoted by  $w(e)$, which is calculated to reflect the significance and connectivity strength between various functions. We are defining paths within this graph as sequences of edges, represented by $\langle u_i, u_{i+1} \rangle$, where $u_0 = s$ and $u_l = t$ demarcate the start and end vertices of the path, respectively. The aggregate length of these paths is the peak of the weights of the constituent edges. Of particular note is the definition of the distance between two vertices, $s$ and $t$, represented as $d(s, t)$, with the shortest paths between them signified by $\sigma_{st}$. Additionally, we are quantifying the number of these shortest paths that traverse through a specific vertex $v$ using $\sigma_{st}(v)$, , being consistent with the notation of Bader et al \cite{1690659}.\\

\textbf{Degree centrality:} We are measuring the degree centrality of a vertex \( v \), denoted as \( \text{deg}(v) \), to quantify the extent of interactions a node has within its neighborhood. This metric reflects the node's importance based on the number of caller-callee connections it maintains \cite{1690659}. \\

%In a network, a node exhibiting high degree centrality is demonstrating connections to a larger proportion of other nodes, signifying its crucial role in the overall network connectivity. We are expressing this centrality as the fraction of nodes to which a given node is connected.

\textbf{The Norm: } By definition of the Euclidean norm of a vector used by velivckovic et al \cite{velivckovic2017graph}  in GATs model, $\mathbf{x} \in \mathbb{R}^n$ as: $    \|\mathbf{x}\|_2 = \sqrt{\sum_{i=1}^{n} x_i^2} $
where $\mathbf{x} = (x_1, x_2, \ldots, x_n)$ represents a vector in an $n$-dimensional real space, and $x_i$ corresponds to the $i$-th element of the vector.

\textbf{Closeness centrality:} is measuring the degree of proximity of a node to all other nodes in the graph, rooted in the concept of distance. For any given node $n$ its closeness centrality is being calculated as the average length of the shortest path from $n$ to every other node in the graph. A node with a higher closeness centrality is more centrally located in the network, thus signifying its greater importance or influence within the network's structure. This metric is being determined by the inverse of the sum of the shortest distances from the node in question to all other nodes \cite{1690659,nieminen1973centrality}. $
    CC(v) = 1/ \sum_{u \in V} d(v, u)
$

In the context of our implementation, this is translating to an assessment of how interconnected a specific function is in relation to the rest of the functions within the code.\\

\textbf{Betweenness centrality:} is quantifying how frequently a node appears on the shortest paths between other nodes, thereby acting as a critical bridge within the network \cite{1690659}. In the context of functions and vulnerabilities within software systems, a function with high betweenness centrality is potentially pivotal in the flow of information or processes. Such a function might have the capacity to significantly influence or impact a multitude of other functions. The pairwise dependency $\delta_{st}(v)$, representing the fraction of shortest paths between nodes $s$ and $t$ that pass through node $v$, is defined as follows:
$
    \delta_{st}(v) = \frac{\sigma_{st}(v)}{\sigma_{st}}
$ 

This leads to the formulation of betweenness centrality for a node $v$, where $s, v, t \in V$, as:$
    BC(v) = \sum _{s\neq v \neq t} \delta_{st}(v)
$

Betweenness centrality is a key metric in network analysis, reflecting a node's ability to control the flow of information or resources by acting as a bridge on the shortest paths between nodes. Nodes with high betweenness centrality are essential for network connectivity, highlighting their role in facilitating communication and interactions. These strategically located nodes ensure efficient dissemination of information by being part of numerous shortest paths connecting various pairs of nodes within the network.

\textbf{Connected components:} are being defined as all subgraphs within which any two vertices are connected to each other by paths, while simultaneously not being connected to any additional vertices in the supergraph. \cite{he2017connected} Our research delineates a connected component in a graph as a set of nodes where each node can access all other nodes within the same set, without external connections. This notion is crucial for analyzing network structures, detecting isolated clusters, and examining connectivity. The study further distinguishes between undirected graphs, where a connected component comprises the largest set of nodes interconnected by paths, and directed graphs, which introduce strongly connected components defined by bidirectional paths between all pairs of nodes.

Our approach is focusing on leveraging the concept of connected components to gain insights into the network's architecture. This analysis is pivotal in identifying potential vulnerabilities or areas of improvement within the network, especially in the context of cybersecurity and software engineering. By understanding the formation and interaction of these components, we are able to devise more effective strategies for network optimization and vulnerability mitigation.

\textbf{The clustering coefficient:} for a node $v$, denoted as $C_i$, is currently being assessed to determine the likelihood of connectivity between two randomly chosen neighbors of this node. This measure is indicative of the number of triangles in which the $i$-th node participates, normalized by the maximum possible number of such triangles. We are computing the overall average clustering coefficient by averaging these individual values across all nodes in the graph. $ 
    C_i = 2t_i / k_i(k_i - 1)
$As this average clustering coefficient approaches 1, it is suggesting an increasing completeness of the graph, characterized by a predominant, cohesive component. A higher coefficient is typically indicative of triadic closure, as observed in denser graphs where triangular formations are more prevalent \cite{saramaki2007generalizations}.

The average clustering for a graph is now being calculated as the average of the local clustering coefficients of all nodes: $
    \bar C_i = \frac{1}{n} \sum_{i=1}^{n} C_i
$
Therefore, the average clustering coefficient is providing an overall indication of the degree of clustering within the entire network. This metric reflects how closely nodes within a graph tend to cluster together, offering insights into the network's structural density and connectivity patterns.

We are defining the quantity of triangles $t_i$ around each node $i$, where $C_i = 0$ indicates that the neighbors of a vertex are not connected, and $C_i = 1$ signifies that neighbors are fully connected. This approach allows us to quantify the degree of local node interconnectivity within the graph, contributing to our understanding of the network's overall structure and cohesiveness.\\

\textbf{Degree Assortativity Coefficient:} We are currently measuring the degree assortativity coefficient, which quantifies the tendency of nodes in a network to connect with other nodes that possess a similar degree. This metric is providing insights into the network's tendency towards assortative or disassortative mixing \cite{newman2003mixing}. Specifically, it determines whether nodes with a high number of connections (high degree) are more likely to connect with other highly connected nodes, or with nodes having fewer connections. The coefficient varies from $-1$ to $1$, where values close to $1$ indicate assortative mixing, implying high-degree nodes predominantly connect with other high-degree nodes. Conversely, values close to $-1$ signify a disassortative mixture, where high-degree nodes are more likely to connect with low-degree nodes. A value around 0 indicates no particular preference in the network's connectivity pattern.

The mathematical definition of the Degree Assortativity Coefficient $r$ is formulated as: $
r = {\sum_{jk} jk (e_{jk} - q_j q_k)}/{\sigma_q^2}.
$
In this formula, $e_{jk}$ represents the fraction of edges in the network that connect a node of degree $j$ to a node of degree $k$. The term $q_j$ denotes the distribution of the remaining degrees of the nodes at the end of an edge, when one end is attached to a node of degree $j$. Finally, $\sigma_q^2$ is the variance of the distribution $q$. This coefficient is playing a crucial role in understanding the structural tendencies of the network, providing valuable insights into how nodes within the network preferentially form connections based on their degrees.\\

\textbf{Cyclomatic complexity} is currently serving as a quantitative measure for evaluating the number of linearly independent paths within a code, thus providing an estimate of the program's complexity \cite{ebert2016cyclomatic}. This metric is instrumental in understanding the intricacy and structural complexity of a program \cite{sarwar2013cyclomatic}. Adapting McCabe's original definition \cite{mccabe1976complexity}, in graph theory terms, the cyclomatic complexity $V(G)$ of a control flow graph $G$ is being defined as follows: $
    V(G) = E - N + 2P
$
In this equation, $E$ represents the number of edges in the graph, $N$ signifies the number of nodes, and $P$ stands for the number of connected components within the graph. This complexity measure is providing a crucial insight into the potential paths and decision points in a program's structure. By quantifying the complexity, we are gaining a deeper understanding of the software's maintainability and potential areas for refactoring. Cyclomatic complexity, therefore, is not just a theoretical construct but a practical tool for guiding the development and maintenance of robust and efficient software systems.

\section{Dataset Results and Interpretation}

This precursory study analyzes three distinct applications from the Python and Java Source languages leveraging open source packages with control flow graphs of: the respective base application, the broken application post-upgrade, and the non-broken application post-upgrade versions, as applicable. For each case, a specific number of functions have been marked as critical, in the CodeQL analysis has marked the specific changes to these functions as the reasons for causing the application to no longer compile. The objective is to gain insights into how the structural and functional attributes of the functions in the graph affect the likelihood of a code breakage, especially in response to modifications and the introduction of new features or patches intended to fix vulnerabilities.

\textbf{CASE 1:} (Table \ref{tab:Case1}) We analyzed an application code base comprising 9626 functions, among which 27 functions were identified as critical, meaning the remediation upgrade caused changes in these functions that would cause the code to break. Upon examination of the resolved code, an increase in the number of nodes and a rise in the average degree were observed following the update. The density of the network remained constant, which aligns with expectations. There was a increasing in the number of connected components, and the degree assortativity coefficient stayed negative and constant. Comparing both versions of the upgrades, the degree assortativity coefficient for the Non Broken Upgrade was much lower than the metric from the Broken Upgrade, while the closeness centrality was higher for Non Broken Upgrades.  

\begin{table*}[h]
\centering
\begin{adjustbox}{width=\textwidth,center}
\begin{tabular}{|l|l|l|l|l|l|}
\hline
 & \textbf{Base} & \multicolumn{2}{c|}{\textbf{Non Broken Upgrade}} & \multicolumn{2}{c|}{\textbf{Broken Upgrade}} \\ \hline
 Number of unique functions & 9621 & \multicolumn{2}{c|}{9621} & \multicolumn{2}{c|}{9621} \\ \hline
 \textbf{nodes kind}&  & \textbf{Unchanged } & \textbf{Changed } & \textbf{Unchanged } & \textbf{Changed } \\ \hline
Number of nodes & 9621 & 9614 & 19 & 9613 & 23 \\ \hline
Number of edges & 15186 & 15178 & 19 & 15176 & 24 \\ \hline
Average degree & 3.1568 & 3.1575 & 2.0 & 3.1574 & 2.087 \\ \hline
Density & 0.0003281 & 0.0003285 & 0.1111 & 0.0003285 & 0.09486 \\ \hline
Num. of connected components & 327 & 327 & 2 & 327 & 2 \\ \hline
Average clustering & 0.06809 & 0.06815 & 0.0 & 0.06815 & 0.0 \\ \hline
Degree assortativity coefficient & -0.08515 & -0.08519 & -0.3996 & -0.08521 & -0.3464 \\ \hline
Avg. betweenness centrality & 0.00036 & 0.000361 &1.63973 & 0.00036 & 2.13699 \\ \hline
Avg. closeness centrality & 0.15857 & 0.15864 & 0.11057 & 0.15871 & 0.0909 \\ \hline
Cyclomatic Complexity & 6219 & 6218 & 4 & 6217 & 5 \\ \hline
\end{tabular}
\end{adjustbox}
\caption{(Case1) Comparative analysis of code metrics from code base, non-broken update and broken upgrade.}
\label{tab:Case1}
\end{table*}

\textbf{CASE 2:} (Table \ref{table:case2})We analyzed an application code base comprising 19,569 functions with 37,615 caller-recipient interactions, among which 3 functions were identified as critical in their failed update, causing the code to break. Upon examination of the metrics, the number of functions affected for both upgrades were significantly higher, however the number of functions remained the same. There was a decrease in the cyclomatic complexity after both upgrades,  while the number of connected components rose. Comparing both versions of the upgrade the degree assortativity coefficient for the Non Broken Upgrade was again lower than the metric from the Broken Upgrade, along with the density of the changed nodes, while the closeness centrality was again higher for Non Broken Upgrades. 

\begin{table*}[h]
\centering
\begin{adjustbox}{width=\textwidth,center}
\begin{tabular}{|l|l|l|l|l|l|}
\hline
\textbf{}         & \textbf{Base} & \multicolumn{2}{l|}{\textbf{Non Broken Upgrade}} & \multicolumn{2}{l|}{\textbf{Broken Upgrade}} \\ \hline
Number of unique functions & 19569 & \multicolumn{2}{l|}{19569} & \multicolumn{2}{l|}{19569} \\ \hline

 \textbf{nodes kind}&  & \textbf{Unchanged } & \textbf{Changed } & \textbf{Unchanged } & \textbf{Changed } \\ \hline

Number of nodes  & 19569         & 17635                        & 2629                       & 17255                             & 3338                            \\ \hline
Number of edges  & 37615         & 33180                        & 4586                       & 32234                             & 5850                            \\ \hline
Average degree   & 3.8443        & 3.7630                       & 3.4888                     & 3.7362                            & 3.5051                          \\ \hline
Density          & 0.00019646    & 0.00021339                   & 0.00132754                 & 0.00021654                        & 0.00105037                      \\ \hline
Number of connected components & 440 & 426                        & 80                         & 429                               & 97                              \\ \hline
Average clustering & 0.05552     & 0.05505                      & 0.04415                    & 0.05467                           & 0.04482                         \\ \hline
Degree assortativity coefficient & -0.07772 & -0.08047        & -0.12601                    & -0.07853                          & -0.12377                        \\ \hline
Avg. betweenness centrality & 0.00018 & 0.00019 &0.000124 & 0.00019 &0.00012 \\ \hline
Avg. closeness centrality & 0.17790 & 0.17686 & 0.18617 & 0.18716 & 0.17640\\ \hline

Cyclomatic Complexity & 18926    & 16397                        & 2117                       & 15837                             & 2706                            \\ \hline
\end{tabular}
\end{adjustbox}
\caption{(CASE 2) Comparative analysis of code metrics from code base, non-broken upgrade and broken upgrade.}
\label{table:case2}
\end{table*}

\textbf{CASE 3:} (Table \ref{table:case3}) We analyzed an application code base comprising 15908 functions with 29449 caller-recipient interactions, among which 6 functions of the upgrade were critical. Upon examination of the metrics, there was a decrease in the cyclomatic complexity after both upgrades,  while the number of connected components again rose. Comparing both versions of the upgrade the Degree assortativity coefficient for the Non Broken Upgrade is this time higher than the metric from the Broken Upgrade, however the density remains lower in the changed nodes, while the closeness centrality was higher for Non Broken Upgrades.

\begin{table*}[h]
\centering
\begin{adjustbox}{width=1\textwidth}
\begin{tabular}{|l|l|l|l|l|l|}
\hline
\textbf{}         & \textbf{Base} & \multicolumn{2}{l|}{\textbf{Non Broken Upgrade}} & \multicolumn{2}{l|}{\textbf{Broken Upgrade}} \\ \hline
Number of unique functions & 15908 & \multicolumn{2}{l|}{15908} & \multicolumn{2}{l|}{15908} \\ \hline

 \textbf{nodes kind}&  & \textbf{Unchanged } & \textbf{Changed } & \textbf{Unchanged } & \textbf{Changed } \\ \hline
 
Number of nodes  & 15908         & 15553                        & 603                       & 15127                             & 1015                            \\ \hline
Number of edges  & 29449         & 28616                        & 912                       & 27695                             & 1816                            \\ \hline
Average degree   & 3.7024        & 3.6798                       & 3.0249                    & 3.6617                            & 3.5783                          \\ \hline
Density          & 0.00023275    & 0.00023661                   & 0.00502471                & 0.00024208                        & 0.00352892                      \\ \hline
Number of connected components & 337 & 341                       & 19                         & 335                               & 28                              \\ \hline
Average clustering & 0.05004     & 0.04834                      & 0.06402                   & 0.04688                           & 0.05890                         \\ \hline
Degree assortativity coefficient & -0.10958 & -0.11123        & -0.10286                   & -0.11203                          & -0.13705                        \\ \hline
Avg. betweenness centrality & 0.00022 & 0.00022 & 0.00018 & 0.00022 & 0.00022 \\ \hline
Avg. closeness centrality & 0.18028 &  0.18032 & 0.17876 & 0.18049 & 0.17684 \\ \hline
Cyclomatic Complexity & 14215    & 13745                        & 347                       & 13238                             & 857                             \\ \hline
\end{tabular}
\end{adjustbox}
\caption{(CASE 3) Comparative analysis of code metrics from code base, non-broken update and broken upgrade}
\label{table:case3}
\end{table*}

\textbf{Statistical Differences between Base and Subgraphs:} Given the consistent direction of difference between the closeness centrality for the three cases above, we decided to further explore those differences. The above analysis that conducting a package upgrade generates two subgraphs of the control flow graph, the functions affected by the upgrade and the functions not affected by the upgrade. We posit that the Broken package upgrade subgraphs are statistically different from a randomly drawn subgraph, and tested this using the closeness centrality value for each node.  We used a T-test an the Kolmogorov-Smirnov (K-S) tests to evaluate differences in means and the distributions, respectively. For Broken Case 1, we observe a T-statistic of -4.881 with a p-value of 0.0012 and a K-S statistic of 0.6916 and a p-value of 8.16e-05, indicating a pronounced deviation in closeness centrality between changed and all nodes. In contrast, Broken Case 2 has a a T-statistic of 8.399 with a p-value of 6.47e-17 and a lower K-S statistic of 0.1161 with a p-value of 1.38e-27,suggesting less variation but still significant differences. Broken Case 3 further corroborates the presence of significant differences with a K-S statistic of 0.2176 and a p-value of 9.81e-34, complemented by a T-statistic of -2.349 and a p-value of 0.019. In summary, the mean values for closeness centrality and the distribution of the closeness centrality values for the nodes appear to be statistically different between subgraphs of changes resulting from  upgrades that break functionality and the base repositories.

The results from the three cases analyzed here exhibit a trend where cyclomatic complexity tends to keep or decrease as upgrades are made to package. This observation suggests a relationship between the resolution of vulnerabilities and the simplification of code structure that may justify further exploration. This ongoing analysis is crucial for understanding the dynamic nature of package vulnerabilities and their impact on overall code complexity.

\begin{figure}[h]
    \centering
    % División izquierda para la tabla
    \begin{minipage}{0.5\textwidth}
        \centering
        \begin{tabular}{|l|c|c|c|}
            \hline
            \textbf{} & \textbf{Case 1} & \textbf{Case 2} & \textbf{Case 3} \\ \hline
            NC & 27 & 3 & 6 \\ \hline
            MSC & 0.5295 & 0.3662 &  0.5045 \\ \hline
            mSC & 0.2461 & 0.1254 &  0.1260 \\ \hline
            ASC & 0.4359 & 0.2185 & 0.3228 \\ \hline
            AGS & 0.4287 & 0.3785 & 0.3153 \\ \hline
        \end{tabular}
    \end{minipage}%
    % División derecha para el gráfico
    \begin{minipage}{0.5\textwidth}
        \centering
        \includegraphics[width=\textwidth]{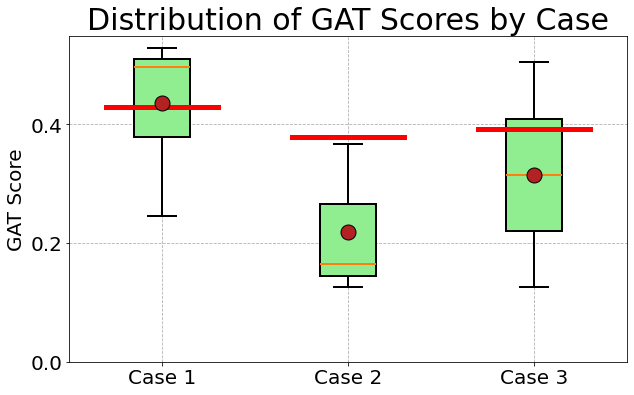}
        \label{fig:gat_score_visualization}
    \end{minipage}
   
\caption{NC: number of Critical Function, MSFC:Max Score Critical, mSFC:Min Score Critical, ASC:Avr Score Critical, and AG:Avr Gat Score for all nodes(Red line)}
\label{tab:modified_gat_results} 
\end{figure}

\textbf{Modified GAT Results} The next step of the analysis attempts to analyze the interconnectedness of the critical functions causing the package upgrades to fail. When we apply the modified GAT model Fig. \ref{tab:modified_gat_results}, the scores obtained. It is necessary to see each case as a specific case. For this, an average GAT score was obtained for each case. Having a high GAT score above average indicates that the vulnerabilities are more critical and necessary since the network depends more on them.

\begin{figure}
    \centerline{\includegraphics[width=1.1\textwidth]{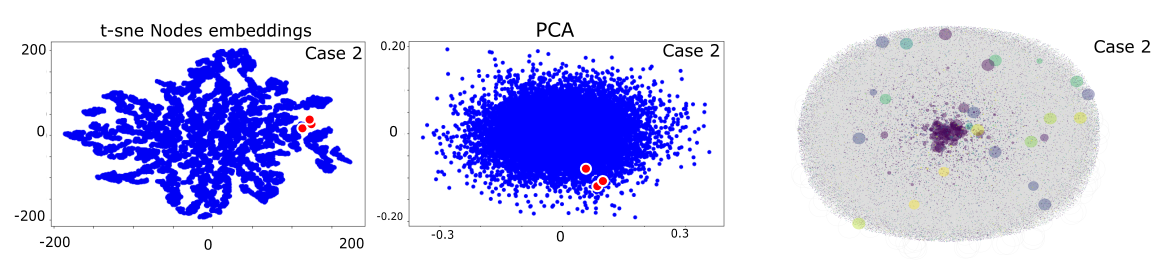}}
    \caption{Visualization of communities for Case 2, using t-SNE and PCA applied to a modified Graph Attention Network (GAT) as usually worked in GAT data analysis,The red dot is the representation of the vulnerability in their respective spaces. The third figure is a graphical representation of the communities within the graph, each distinguished by a unique color. }
    \label{fig:pca}
\end{figure}

Principal Component Analysis (PCA) and t-Distributed Stochastic Neighbor Embedding (t-SNE) \cite{velivckovic2017graph} are dimensionality reduction techniques applied alongside GATs to enhance visualization and interpretability as we do in the Case 2 Fig.\ref{fig:pca}. While PCA projects data into a lower-dimensional space preserving variance, t-SNE focuses on maintaining local relationships, making it particularly useful for visualizing high-dimensional data generated by GATs in a way that highlights patterns and relationships within the graph structure.

The enhanced GAT Score significantly augments our ability to discern the connectivity and importance of critical functions within a system's context. By employing a normalized approach, where scores closer to 1 denote higher importance, we gain nuanced insights into the integral nature of specific vulnerabilities. This is exemplified in the data for CASE 1, with an average GAT score for the entire graph of 0.4287, and most of the 27 critical functions are above this value (see Fig. \ref{tab:modified_gat_results}). In CASE 2, the GAT scores for all critical functions are less than the average, with a maximum GAT score for a critical function of 0.3662, while the average GAT Score for the entire graph is 0.3785. This implies that the criticality of these functions is lower. It's noteworthy that these scores are not solely reliant on the degree of connectivity; rather, they integrate a weighted combination of degree, norm, and centrality metrics. This holistic approach allows us to identify functions that, despite having a lower degree of connectivity, hold substantial significance within the network's overall architecture. Such revelations underscore the complexity of network dynamics and the crucial role of advanced analytical tools in unveiling the intricate interplay of functions within a software system.

\begin{figure}
    \centering
    \includegraphics[width=1\linewidth]{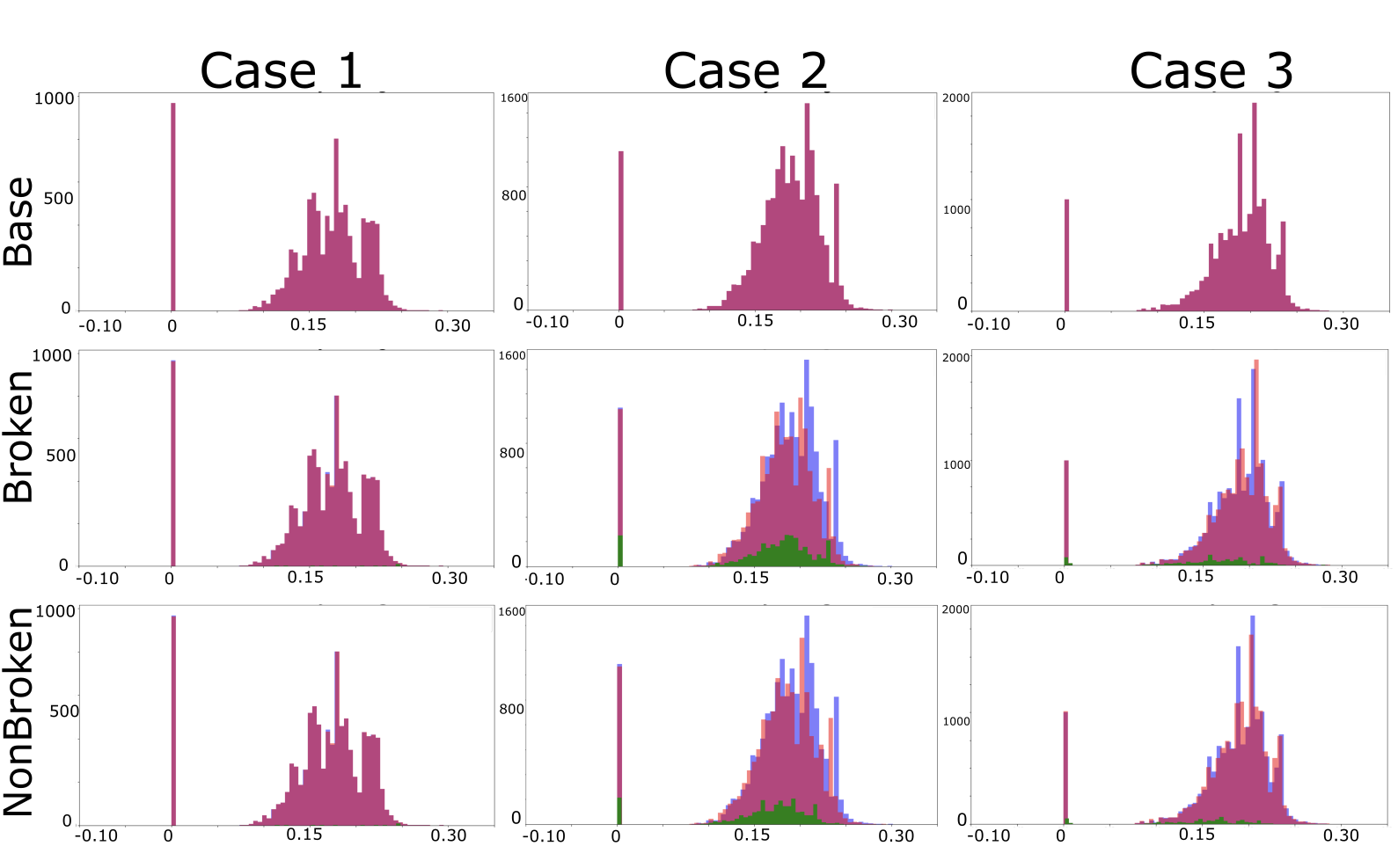}
    \caption{Closeness Centrality Histogram, y-axes counts and x-axes Closeness Centrality. Cases 1, 2, 3, and 4 with their respective base code, broken upgrade, and non-broken upgrade as applicable. Light blue indicates the centrality of all nodes, light red represents unchanged nodes, and green denotes changed nodes.}
    \label{fig:clustering}
\end{figure}

\begin{figure}
    \centerline{\includegraphics[width=1.1\textwidth]{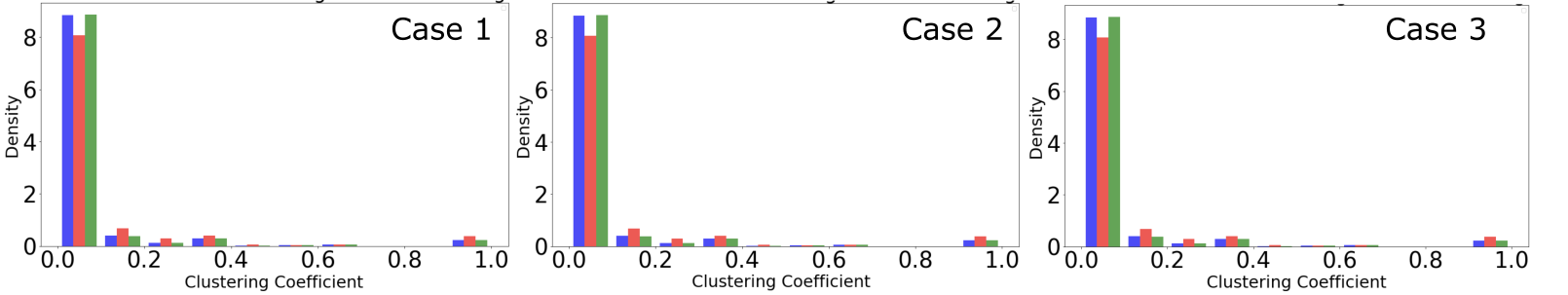}}
    \caption{ Normalized cases of clustering coefficient histogram. Blue: All nodes, Red: Changed nodes, Green: Non changed nodes.}
    \label{fig:clustering+}
\end{figure}

\section{Discussion}
The results of our extensive analysis of several code bases reveal an intricate dynamic between package vulnerabilities and the complexity of the code. This study contributes to the current discourse in package development by highlighting the nuanced relationship between the vulnerabilities after the publication and evolution of the code.This requires not only immediate code updates, but also a deeper understanding of the underlying dynamics of these vulnerabilities. 

Our analysis of CASES 1, 2, and 3 has led to several key observations about the impact of vulnerability resolution on code structure. We noticed a trend of decreased or consistent total cyclomatic complexity when comparing base cases with broken and intact vulnerability resolutions. This pattern indicates that effectively addressing vulnerabilities tends to simplify the code structure, aligning with best practices for developing maintainable and less error-prone packages. However, managing the complexity associated with package vulnerabilities remains a formidable challenge. In particular, we observed an increase in the number of connected components from the base case to the non-broken update, suggesting that remediation efforts not only resolve vulnerabilities but also improve the connectivity and robustness of the code architecture. Despite this, even minor vulnerabilities can have far-reaching effects across the entire application, underscoring the need for a comprehensive approach to understanding and mitigating the impacts of package updates. The intricate nature of this challenge is further highlighted in cross-centrality graphs (\ref{fig:clustering}, \ref{fig:clustering+}), where significant variations in centrality between nodes highlight the complex interdependencies within the code. These findings emphasize the critical importance of detailed graph-based analyzes for navigating the complexities of software vulnerability management and improvement.

In addition, our analysis sheds light on the connectivity patterns of the function within the code bases. The coefficient of negative degree supply consistently observed in multiple studies indicates that the code functions tend to connect with a diverse variety of other functions, instead of predominantly linked to similar functionalities. This diversity in connectivity patterns has deep implications to understand how vulnerabilities could spread through a code base and affect their general integrity.

This ambiguity highlights a critical gap in current understanding and requires more research. It emphasizes the importance of a strategic approach to code updates, where essential functionalities such as configuration are refined and maintained, instead of being completely eliminated.

The significance of the modified GAT score is further underscored by our observations of the wide dispersion of nodes and low density within the graphs analyzed. This dispersion necessitates a nuanced approach to understanding the role of each node within the network, focusing not only on the number of connections it can generate but also on the number of paths that traverse through it. The modified GAT model offers a normalized view of how interconnected a function is in relation to its degree, norm, and closeness centrality metrics. This comprehensive perspective is crucial for effective vulnerability management and software maintenance, as it highlights the critical nodes that warrant prioritized attention. By integrating the insights from the GAT score with observations of node propagation and density, stakeholders are equipped with a powerful tool to identify and address the most significant vulnerabilities, optimizing resource allocation and enhancing the security and reliability of software systems. This blend of quantitative measures and attention-based analysis signifies a significant advancement in our ability to strategically mitigate vulnerabilities and strengthen software infrastructure.

Our study underscores the dynamic nature of package vulnerabilities and their impact on code complexity. It calls for continued vigilance and strategic intervention in the package development lifecycle, stressing the need for robust monitoring and adaptive strategies to manage the evolving landscape of package vulnerabilities.

\section{Conclusion}
This paper embarks on a comprehensive exploration, leveraging the intricate capabilities of knowledge graphs to delve into the dynamics of opensource package function networks. Central to our investigation is the identification and analysis of vulnerable functions, where we scrutinize their interactions and assess the impact of their mitigation on the codebase. Our research reveals a notable insight: targeted remediation of specific vulnerabilities tends to preserve the overall network's connectivity, underscoring the package structure's inherent resilience.

During our analysis, we noted a pattern of decreasing vulnerabilities following successive package updates, prompting a pivotal inquiry: are these vulnerabilities conclusively resolved, or do they subtly embed themselves into subsequent features? This ambiguity signals a compelling need for further work into the lifecycle of vulnerabilities within package development, promising to enrich the discourse in this field significantly.

At the heart of our methodology is the odified Graph Attention Network (GAT), especially its attention mechanism. Through the integration of node-centric metrics—such as degree centrality, norm, and closeness centrality—our approach refines the network's ability to discern detailed aspects of the graph's architecture and the nuances of node characteristics. This methodological advancement facilitates a nuanced portrayal of the network, yielding a comprehensive understanding of node interrelations and their significance.

Furthermore, our investigation brings to light the existence of latent vulnerabilities within the most critical segments of the code, initially perceived as flawless. These covert vulnerabilities represent significant security risks, with the potential to compromise vital components, including databases and core functionalities. To address these issues, we advocate for an in-depth and ongoing code analysis from its inception. Utilizing knowledge graphs as both historical and dynamic monitoring tools enables proactive surveillance of vulnerabilities.

For future works, our research direction will focus on methodological improvements, particularly in dissecting the interconnectivity between functions. The number of case analyzed was small with significant complexity, necessitating additional application repositories for comprehensive analysis to draw more robust conclusions. By evaluating various aspects such as variable types, execution times, and functional dependencies, we aim to unravel the importance of specific functions within the network. This comprehensive strategy is designed to offer deeper insights into the structural integrity and vulnerabilities of software systems, thereby making a substantial contribution to enhancing package security and dependability. Our endeavors are geared towards the development of robust software systems capable of navigating the complexities of contemporary cyber threats, marking a significant stride forward in the realm of package vulnerability analysis and cybersecurity.

\section*{Acknowledgment}
This research was supported in part by NSF grant CCF-2109988.

%\clearpage
\bibliographystyle{plain}
\bibliography{vul01,knowledgeGraph,topology,ciberSecurity,related}

\begin{thebibliography}{10}

\bibitem{alanazi2022scada}
Manar Alanazi, Abdun Mahmood, and Mohammad Jabed~Morshed Chowdhury.
\newblock Scada vulnerabilities and attacks: A review of the state-of-the-art
  and open issues.
\newblock {\em Computers \& Security}, page 103028, 2022.

\bibitem{alfadel2023empirical}
Mahmoud Alfadel, Diego~Elias Costa, and Emad Shihab.
\newblock Empirical analysis of security vulnerabilities in python packages.
\newblock {\em Empirical Software Engineering}, 28(3):59, 2023.

\bibitem{9802103}
Raghavendra~Rao Althar, Debabrata Samanta, Manjit Kaur, Dilbag Singh, and
  Heung-No Lee.
\newblock Automated risk management based software security vulnerabilities
  management.
\newblock {\em IEEE Access}, 10:90597--90608, 2022.

\bibitem{arulselvan2009detecting}
Ashwin Arulselvan, Clayton~W Commander, Lily Elefteriadou, and Panos~M
  Pardalos.
\newblock Detecting critical nodes in sparse graphs.
\newblock {\em Computers \& Operations Research}, 36(7):2193--2200, 2009.

\bibitem{1690659}
David~A. Bader and Kamesh Madduri.
\newblock Parallel algorithms for evaluating centrality indices in real-world
  networks.
\newblock In {\em 2006 International Conference on Parallel Processing
  (ICPP'06)}, pages 539--550, 2006.

\bibitem{ebert2016cyclomatic}
Christof Ebert, James Cain, Giuliano Antoniol, Steve Counsell, and Phillip
  Laplante.
\newblock Cyclomatic complexity.
\newblock {\em IEEE software}, 33(6):27--29, 2016.

\bibitem{10.1145/3379597.3387501}
Jiahao Fan, Yi~Li, Shaohua Wang, and Tien~N. Nguyen.
\newblock A c/c++ code vulnerability dataset with code changes and cve
  summaries.
\newblock In {\em Proceedings of the 17th International Conference on Mining
  Software Repositories}, MSR '20, page 508–512, New York, NY, USA, 2020.
  Association for Computing Machinery.

\bibitem{giray2021software}
G{\"o}rkem Giray.
\newblock A software engineering perspective on engineering machine learning
  systems: State of the art and challenges.
\newblock {\em Journal of Systems and Software}, 180:111031, 2021.

\bibitem{goseva2015capability}
Katerina Goseva-Popstojanova and Andrei Perhinschi.
\newblock On the capability of static code analysis to detect security
  vulnerabilities.
\newblock {\em Information and Software Technology}, 68:18--33, 2015.

\bibitem{he2017connected}
Lifeng He, Xiwei Ren, Qihang Gao, Xiao Zhao, Bin Yao, and Yuyan Chao.
\newblock The connected-component labeling problem: A review of
  state-of-the-art algorithms.
\newblock {\em Pattern Recognition}, 70:25--43, 2017.

\bibitem{Hogan_2021}
Aidan Hogan, Eva Blomqvist, Michael Cochez, Claudia D'amato, Gerard~De Melo,
  Claudio Gutierrez, Sabrina Kirrane, Jos{\'{e} } Emilio~Labra Gayo, Roberto
  Navigli, Sebastian Neumaier, Axel-Cyrille~Ngonga Ngomo, Axel Polleres,
  Sabbir~M. Rashid, Anisa Rula, Lukas Schmelzeisen, Juan Sequeda, Steffen
  Staab, and Antoine Zimmermann.
\newblock Knowledge graphs.
\newblock {\em {ACM} Computing Surveys}, 54(4):1--37, jul 2021.

\bibitem{imtiaz2021comparative}
Nasif Imtiaz, Seaver Thorn, and Laurie Williams.
\newblock A comparative study of vulnerability reporting by software
  composition analysis tools.
\newblock In {\em Proceedings of the 15th ACM/IEEE International Symposium on
  Empirical Software Engineering and Measurement (ESEM)}, pages 1--11, 2021.

\bibitem{jia2018practical}
Yan Jia, Yulu Qi, Huaijun Shang, Rong Jiang, and Aiping Li.
\newblock A practical approach to constructing a knowledge graph for
  cybersecurity.
\newblock {\em Engineering}, 4(1):53--60, 2018.

\bibitem{liu2022recent}
Kai Liu, Fei Wang, Zhaoyun Ding, Sheng Liang, Zhengfei Yu, and Yun Zhou.
\newblock Recent progress of using knowledge graph for cybersecurity.
\newblock {\em Electronics}, 11(15):2287, 2022.

\bibitem{liu2022review}
Kai Liu, Fei Wang, Zhaoyun Ding, Sheng Liang, Zhengfei Yu, and Yun Zhou.
\newblock A review of knowledge graph application scenarios in cyber security.
\newblock {\em arXiv preprint arXiv:2204.04769}, 2022.

\bibitem{mccabe1976complexity}
Thomas~J McCabe.
\newblock A complexity measure.
\newblock {\em IEEE Transactions on software Engineering}, (4):308--320, 1976.

\bibitem{newman2003mixing}
Mark~EJ Newman.
\newblock Mixing patterns in networks.
\newblock {\em Physical review E}, 67(2):026126, 2003.

\bibitem{nieminen1973centrality}
UJ~Nieminen.
\newblock On the centrality in a directed graph.
\newblock {\em Social science research}, 2(4):371--378, 1973.

\bibitem{park2019estimating}
Namyong Park, Andrey Kan, Xin~Luna Dong, Tong Zhao, and Christos Faloutsos.
\newblock Estimating node importance in knowledge graphs using graph neural
  networks.
\newblock In {\em Proceedings of the 25th ACM SIGKDD international conference
  on knowledge discovery \& data mining}, pages 596--606, 2019.

\bibitem{platt2019network}
Edward~L Platt.
\newblock {\em Network science with Python and NetworkX quick start guide:
  explore and visualize network data effectively}.
\newblock Packt Publishing Ltd, 2019.

\bibitem{rodriguez2022arachne}
Oliver~Alvarado Rodriguez, Zhihui Du, Joseph Patchett, Fuhuan Li, and David~A
  Bader.
\newblock Arachne: An arkouda package for large-scale graph analytics.
\newblock In {\em 2022 IEEE High Performance Extreme Computing Conference
  (HPEC)}, pages 1--7. IEEE, 2022.

\bibitem{saramaki2007generalizations}
Jari Saram{\"a}ki, Mikko Kivel{\"a}, Jukka-Pekka Onnela, Kimmo Kaski, and Janos
  Kertesz.
\newblock Generalizations of the clustering coefficient to weighted complex
  networks.
\newblock {\em Physical Review E}, 75(2):027105, 2007.

\bibitem{sarwar2013cyclomatic}
Mir Muhammd~Suleman Sarwar, Sara Shahzad, and Ibrar Ahmad.
\newblock Cyclomatic complexity: The nesting problem.
\newblock In {\em Eighth International Conference on Digital Information
  Management (ICDIM 2013)}, pages 274--279. IEEE, 2013.

\bibitem{seacord_2014}
Robert Seacord.
\newblock Secure coding to prevent vulnerabilities.
\newblock Carnegie Mellon University, Software Engineering Institute's Insights
  (blog), May 2014.
\newblock Accessed: 2023-Nov-16.

\bibitem{sikos2023cybersecurity}
Leslie~F Sikos.
\newblock Cybersecurity knowledge graphs.
\newblock {\em Knowledge and Information Systems}, pages 1--21, 2023.

\bibitem{sparks2007automated}
Sherri Sparks, Shawn Embleton, Ryan Cunningham, and Cliff Zou.
\newblock Automated vulnerability analysis: Leveraging control flow for
  evolutionary input crafting.
\newblock In {\em Twenty-Third Annual Computer Security Applications Conference
  (ACSAC 2007)}, pages 477--486. IEEE, 2007.

\bibitem{szabo2023incrementalizing}
Tam{\'a}s Szab{\'o}.
\newblock Incrementalizing production codeql analyses.
\newblock {\em arXiv preprint arXiv:2308.09660}, 2023.

\bibitem{varela2023feature}
{\'A}ngel~Jes{\'u}s Varela-Vaca, Diana Borrego, Mar{\'\i}a~Teresa
  G{\'o}mez-L{\'o}pez, Rafael~M Gasca, and A~German M{\'a}rquez.
\newblock Feature models to boost the vulnerability management process.
\newblock {\em Journal of Systems and Software}, 195:111541, 2023.

\bibitem{velivckovic2017graph}
Petar Veli{\v{c}}kovi{\'c}, Guillem Cucurull, Arantxa Casanova, Adriana Romero,
  Pietro Lio, and Yoshua Bengio.
\newblock Graph attention networks.
\newblock {\em arXiv preprint arXiv:1710.10903}, 2017.

\bibitem{xia2023empirical}
Boming Xia, Tingting Bi, Zhenchang Xing, Qinghua Lu, and Liming Zhu.
\newblock An empirical study on software bill of materials: Where we stand and
  the road ahead.
\newblock {\em arXiv preprint arXiv:2301.05362}, 2023.

\bibitem{xia2023coupled}
Feng Xia, Xin Chen, Shuo Yu, Mingliang Hou, Mujie Liu, and Linlin You.
\newblock Coupled attention networks for multivariate time series anomaly
  detection.
\newblock {\em IEEE Transactions on Emerging Topics in Computing}, 2023.

\bibitem{yan2020review}
Zhihao Yan and Jingju Liu.
\newblock A review on application of knowledge graph in cybersecurity.
\newblock In {\em 2020 International Signal Processing, Communications and
  Engineering Management Conference (ISPCEM)}, pages 240--243. IEEE, 2020.

\bibitem{zhu2021constructing}
Kailong Zhu, Yuliang Lu, Hui Huang, Lu~Yu, and Jiazhen Zhao.
\newblock Constructing more complete control flow graphs utilizing directed
  gray-box fuzzing.
\newblock {\em Applied Sciences}, 11(3):1351, 2021.

\bibitem{zomorodian2009computational}
Afra Zomorodian.
\newblock Computational topology.
\newblock {\em Algorithms and theory of computation handbook}, 2(3), 2009.

\end{thebibliography}

\end{document}